\newif\ifllncs
\llncsfalse
\newif\ifiacrtrans
\iacrtranstrue

\ifllncs
\documentclass[runningheads,natbib=true]{llncs}
\fi
\ifiacrtrans
\documentclass[]{iacrtrans}
\fi
\usepackage[acronym]{glossaries} 
\usepackage{pifont}
\usepackage{fontawesome5}
\usepackage{booktabs}
\newcommand{\cmark}{\ding{51}}
\newcommand{\xmark}{\ding{55}}

\newacronym{rs}{RecSys}{Recommender System}
\newacronym{prs}{PrivRecSys}{Privacy-Aware Recommender System}
\newacronym{rg}{RG}{Recommendation Generator}
\newacronym{rp}{RP}{Recommendation Provider}
\newacronym{rd}{RD}{Recommendation Dataset}
\newacronym{ps}{PS}{Prediction Sets}
\newacronym{rm}{RecModel}{Recommendation Model}
\newacronym{rds}{RecDS}{Recommendation Dataset}

\newacronym{soc}{SOC}{System on Chips}
\newacronym{p2p}{P2P}{Peer-to-Peer}

\newacronym{dp}{DP}{Differential Privacy}
\newacronym{ldp}{LDP}{Local Differential Privacy}
\newacronym{ml}{ML}{Machine Learning}
\newacronym{gdpr}{GDPR}{General Data Protection Regulation}
\newacronym{ccpa}{CCPA}{California Consumer Privacy Act}
\newacronym{he}{HE}{Homomorphic Encryption}
\newacronym{mpc}{MPC}{Multi Party Computation}
\newacronym{2pc}{2-PC}{Two Party Computation}
\newacronym{gc}{GC}{Garbled Circuit}
\newacronym{phe}{PHE}{Partial HE}
\newacronym{she}{SHE}{Somewhat HE}
\newacronym{fhe}{FHE}{Fully HE}
\newacronym{pir}{PIR}{Private Information Retrieval}
\newacronym{knn}{KNN}{K Nearest neighbor}
\newacronym{ot}{OT}{Oblivious Transfer}
\newacronym{ttp}{TTP}{Trusted Third Party}
\newacronym{opse}{OPSE}{Order Preserving Searchable Encryption}
\newacronym{fss}{FSS}{Functional Secret Sharing}
\newacronym{orom}{OROM}{Oblivious Read Only Memory}
\newacronym{oram}{ORAM}{Oblivious RAM}
\newacronym{cf}{CF}{Collaborative Filtering}
\newacronym{lwe}{LWE}{Learning With Errors}

\usepackage{graphicx}

\usepackage[colorlinks=true,
            citecolor=blue,
            linkcolor=blue,
            urlcolor=blue,
            pdfstartview=FitH,
            bookmarks=true,
            bookmarksopen=true,
            bookmarksdepth=2,
            ]{hyperref}
\usepackage[capitalize]{cleveref}

\usepackage{todonotes}               
\usepackage{amsmath}
\usepackage{amsfonts}
\authorrunning{Mukherjee, Walch, Meisingseth, Lex, Rechberger}
\titlerunning{Privacy-Aware Recommender System}

\author{Shibam Mukherjee\inst{1,2} \and Roman Walch\inst{1,3} \and Fredrik Meisingseth\inst{1} \and \\Elisabeth Lex\inst{1} \and Christian Rechberger\inst{1}}

\institute{Graz University of Technology (Austria) \\
    \email{firstname.lastname@tugraz.at} \and Know-Center GmbH (Austria) \and TACEO (Austria)}

\begin{document}
\title{Hiding Your Awful Online Choices Made More Efficient and Secure: A New Privacy-Aware Recommender System}
%
%
%
%
%
\maketitle              
\ifiacrtrans
\fi

\begin{abstract}
Recommender systems are an integral part of online platforms that recommend new content to users with similar interests.
However, they demand a considerable amount of user activity data where, if the data is not adequately protected, constitute a critical threat to the user privacy.
Privacy-aware recommender systems enable protection of such sensitive user data while still maintaining a similar recommendation accuracy compared to the traditional non-private recommender systems.
However, at present, the current privacy-aware recommender systems suffer from a significant trade-off between privacy and computational efficiency.
For instance, it is well known that architectures that rely purely on cryptographic primitives offer the most robust privacy guarantees, however, they suffer from substantial computational and network overhead.
Thus, it is crucial to improve this trade-off for better performance.
This paper presents a novel privacy-aware recommender system that combines privacy-aware machine learning algorithms for practical scalability and efficiency with cryptographic primitives like Homomorphic Encryption and Multi-Party Computation - without assumptions like trusted-party or secure hardware - for solid privacy guarantees.
Experiments on standard benchmark datasets show that our approach results in time and memory gains by three orders of magnitude compared to using cryptographic primitives in a standalone for constructing a privacy-aware recommender system.
Furthermore, for the first time our method makes it feasible to compute private recommendations for datasets containing 100 million entries, even on memory-constrained low-power SOC (System on Chip) devices.

\ifllncs
\fi
\end{abstract}

\section{Introduction}

\gls{rs} have become integral parts of our online world, helping users navigate the vast information in the online space by providing them with the most suitable choices. Currently, all major web-based industries, including online health, finance, or entertainment platforms use \gls{rs} to provide a better user experience. Due to the substantial trade-off between privacy and performance, currently most of the \gls{rs} research primarily focuses on building practical algorithms and relevant datasets for increased accuracy and efficiency, thus guaranteeing privacy mostly during data transmission due to its negligible impact on performance.

In addition to not protecting the data-owner's data (when outsourced to a cloud), such \gls{rs} also inherently does not protect the user query and response from the cloud. For example, if the cloud knows the final response, it can also manipulate it. In the past several years, many works have suggested creating various \gls{prs} approaches that improve the privacy-performance trade-off by adopting trust-based security models, such as non-colluding parties~\cite{erkin2012generating,wu2019privacy,zhu2016secure,wong2009secure,servan2022private,kesarwani2018efficient}, secure hardware~\cite{ahmed2020nearest}, among others. Unfortunately, such security assumptions do not guarantee user data protection against colluding parties or attacks against secure hardwares~\cite{nilsson2020survey}, thus leaking the user information to the cloud. Related works like~\cite{wang2020quickn} take a different approach by using \gls{opse} with R-trees structure to construct an efficient \gls{prs} without trust assumptions, however, they do not take into account information leakage through access pattern
~\cite{lambregts2022volume,markatou2019full}. Recent work by Chen et al. SANNS~\cite{sanns} proposes a \gls{prs} construction by solely relying on privacy-preserving cryptographic primitives with strong security assumptions like use of no trusted-party or secure hardware, in addition to providing resistance against access pattern analysis. However, due to the use of privacy-preserving cryptographic primitives like \gls{he} and \gls{mpc}, their proposal is computationally more expensive compared to the other approaches with weaker security assumptions. Refer to \Cref{tab:comparisionbetweenpriorworks} for an overview comparing different \gls{prs} approaches from the last years. Notably, even though crucial, in the previous literature, protection towards the data-owner's data (\gls{rds}) when outsourced to a recommendation cloud was found missing. Previous works like SANNS assumed that the data-owner and the recommendation cloud are one single entity which might not always be the case.

\subsection{Main Contributions}

\noindent \textit{First Scalable Private Implementation.} For the first time, to the best of our knowledge, our work demonstrates a truly scalable \gls{prs}, capable of generating private recommendations from datasets with 100 million entries, pushing the limits of dataset size by \textit{an order of magnitude} in comparison to the earlier state-of-the-art trust based or trust less approaches.

\noindent \textit{Improved Time, Memory and Privacy.} As a direct consequence of combining privacy-aware \gls{ml} with cryptographic primitives like \gls{he} and \gls{mpc}, our experiments show a time and memory improvement by \textit{three orders of magnitude} when computing a single recommendation compared to using only cryptographic primitives. Additionally, our new approach enables the data-owner to protect its \gls{rds} containing historic user data (like ratings) and other Personal Identifiable Information (PII) from the cloud. This allows more user privacy, especially to the vulnerable groups.

\noindent \textit{Portability and Efficiency.} For the first time, to the best of our knowledge, we present a \gls{prs} capable of generating private recommendations for a dataset with 100 million entries under 18 seconds on a standard consumer device, suitable for batched private recommendations. Moreover, we also demonstrate the efficiency of our \gls{prs} by hosting it on a low-power SOC device.

Until our work, SANNS~\cite{sanns} was the only state-of-the-art \gls{prs} requiring no trust assumptions. However, as far as we are aware, there is no public implementation of SANNS at the time of writing this paper. Thus, to further expedite the open-source research in the direction of efficient \gls{prs} implementations, we are making our implementation publicly available.\footnote{\url{https://extgit.iaik.tugraz.at/krypto/privacyawarerecsys}}

\subsection{Preliminaries}

\noindent \textit{Homomorphic Encryption.} \gls{he}
~\cite{brakerski2011efficient,gentry2013homomorphic} enables a server to privately compute on the client's encrypted data without having to decrypt it first. \gls{he} natively allows only basic arithmetic operations (addition and multiplication), but one can construct and evaluate any other function using them. The native operations of \gls{he} are fast, however, when evaluating other functions like comparison, \gls{mpc} outperforms even the most optimized \gls{he}~\cite{chillotti2020tfhe} implementation by several orders of magnitude. More specifically, \gls{he} can be used for comparisons only when it is instantiated over Boolean values. 

\noindent \textit{Multi-Party Computation.} \gls{mpc} allows multiple parties to jointly compute a function over their inputs while keeping their inputs private. This work uses \gls{gc}~\cite{yao1986generate} as the \gls{mpc} protocol where the client and the recommendation cloud are the two parties. Secret sharing is another important \gls{mpc} technique that enables splitting up a secret among multiple parties such that a single secret share reveals nothing about the original secret.

\noindent \textit{Private Information Retrieval.} By observing the read access pattern of a client request, a malicious cloud can infer sensitive information about the client's encrypted data stored on the cloud
\cite{lambregts2022volume,markatou2019full}. \gls{pir}~\cite{chor1995private,crescenzo2000single} protects against such attacks by making read operation indistinguishable from each other, thus hiding the access pattern. This work uses a trivial \gls{mpc}-based \gls{pir} approach discussed later.

\section{Related Works}

One can divide the problem of constructing a \gls{prs} into two main sub parts, namely, data protection and private computation.

\noindent \textit{Data Protection.} Besides the recommendation algorithm, the \gls{rds} plays a crucial role determining the recommendation quality and thus it is in the best interest of any data-owner to protect it. In previous works like SANNS~\cite{sanns}, the authors assumed that the data-owner and the cloud are one entity, however, in our work, we want to argue that for many data-owners, especially small to medium sized, it is infeasible to develop and deploy in-house \gls{prs}. Thus, a majority of them will rely on third-party clouds, requiring them to forfeit their \gls{rds} to the cloud, potentially containing sensitive user/client information. One trivial approach to hide the \gls{rds} is to store it encrypted on the cloud~\cite{wu2019privacy,kesarwani2018efficient,servan2022private,wong2009secure,zhu2016secure,ahmed2020nearest}. However, most of these proposals assume a trust-based model with non-colluding parties or use of secure hardwares like Intel SGX. Interestingly, we observed that one can also partially hide the \gls{rds} information from the cloud by using \gls{cf}~\cite{herlocker2000explaining,mullner2022reuseknn}, which is a nearest neighbor algorithm that leverages similarity metrics like cosine similarity to find the $k$ most similar neighbors (neighbor-based \gls{cf}) for a requesting client and recommend the items from these neighbors to the client. This implies that given a \gls{rds}, one can use \gls{cf} to pre-compute the recommendations for all the possible items present in the \gls{rds} that a client has not seen yet (potential future recommendations) and store them in a smaller and more private reduced-dataset. One can consider the reduced-dataset as a \gls{rm} because whenever a client requests for a recommendation, the cloud uses this to generate recommendations. Additionally, when constructing the \gls{rm} with \gls{cf} algorithm, one gets the flexibility of choosing similar neighbor such that the vulnerable neighbors (users) are never used to generate recommendations. Such a protection, despite being crucial, was found missing in the previous \gls{prs} literature.

\noindent \textit{Private Computation.} Another major challenge of constructing a cloud-based \gls{prs} is to compute recommendations privately without revealing any intermediate or final results, including results inferred from exploits like access pattern analysis. Approaches like~\cite{zhu2016secure,wu2019privacy} do not protect against access pattern analysis, whereas \cite{wang2020quickn,kesarwani2018efficient,servan2022private,ahmed2020nearest} provide such a protection but only with trust assumptions. Purely relying on cryptographic primitives, SANNS~\cite{sanns} is the only \gls{prs} protecting against access pattern analysis without a trusted party. However, due to the use of expensive privacy-preserving cryptographic primitives, namely \gls{he} and \gls{mpc}, SANNS comes with a higher computational and communication overhead in comparison to the other trust based approaches.

\begin{table}[ht]
\setlength{\tabcolsep}{6pt}
  \centering
  \caption{Comparison with related works. (1) No trusted third party, secure hardware, key sharing or the user itself is the data-owner. (2) Protection even against access pattern analysis. (3) Scalable up to at least 10 million dataset entries. (4) Low CPU, memory and network footprint. (\textcolor{darkgray}{\textbf{?}}) Unsure.}
  \label{tab:comparisionbetweenpriorworks}
  \resizebox{\linewidth}{!}{%
  \begin{tabular}{lcccc}
  \toprule
  \multicolumn{1}{c}{Work} & {No Assumptions}$^{1}$ & {Private Computation}$^{2}$ & {Scalable}$^{3}$ & {Low Overhead}$^{4}$  \\ 
  \midrule
    Wong et al. `09~\cite{wong2009secure} & \textcolor{darkgray}{\xmark} & \textcolor{darkgray}{\xmark} & \textcolor{darkgray}{\textbf{\cmark}} & \textcolor{darkgray}{\cmark} \\
    Zhu et al. `16~\cite{zhu2016secure} & \textcolor{darkgray}{\xmark} & \textcolor{darkgray}{\xmark} &  \textcolor{darkgray}{\cmark} & \textcolor{darkgray}{\cmark} \\
    Kesarwani et al. `18~\cite{kesarwani2018efficient} & \textcolor{darkgray}{\xmark} & \textcolor{darkgray}{\cmark} & \textcolor{darkgray}{\xmark} & \textcolor{darkgray}{\xmark} \\
    Wu et al. `19~\cite{wu2019privacy} & \textcolor{darkgray}{\xmark} & \textcolor{darkgray}{\cmark} & \textcolor{darkgray}{\xmark} & \textcolor{darkgray}{\xmark} \\
    Wang et al. `20~\cite{wang2020quickn} & \textcolor{darkgray}{\xmark} & \textcolor{darkgray}{\xmark} & \textcolor{darkgray}{\cmark} & \textcolor{darkgray}{\cmark} \\
    Ahmed et al. `20~\cite{ahmed2020nearest} & \textcolor{darkgray}{\xmark} & \textcolor{darkgray}{\cmark} & \textcolor{darkgray}{\textbf{?}} & \textcolor{darkgray}{\cmark} \\
    Chen et al. `20~\cite{sanns} & \textcolor{darkgray}{\cmark} & \textcolor{darkgray}{\cmark} & \textcolor{darkgray}{\cmark} & \textcolor{darkgray}{\xmark} \\
    Servan-Schreiber et al. `22~\cite{servan2022private} & \textcolor{darkgray}{\xmark} & \textcolor{darkgray}{\textbf{?}} & \textcolor{darkgray}{\cmark} & \textcolor{darkgray}{\cmark}\\ 
    \midrule
    \textbf{Our Approach} & \textcolor{darkgray}{\cmark} & \textcolor{darkgray}{\cmark} & \textcolor{darkgray}{\cmark} & \textcolor{darkgray}{\cmark}\\
  \bottomrule
  \end{tabular}
  }
\end{table}

\section{Private Recommender System}

We begin by describing the threat model (\Cref{subsection:threat_model}) of our approach, followed by a high level description (\Cref{subsection:high_level_protocol}) of our protocol. Finally, a more detailed description of our private recommender system is provided in \Cref{subsec:dataowner_processing} \Cref{subsec:cloud_processing}.

\subsection{Threat Model}
\label{subsection:threat_model}

\noindent \textit{Client.} Our model protects the client query and the recommendations both from the data-owner and the cloud without any trust assumptions, including protection against access pattern analysis. 
For generating the \gls{rm} and providing the optional recommendation feedbacks, the client has to forfeit its sensitive historic/feedback data (PII) like personal preference (ratings) and usage patterns to the data-owner.\footnote{\Cref{sec:furtherdiscussions} explores possible strategies to minimize the historic preference leakage.}
Furthermore, when generating the \gls{rm}s, a vulnerable client can request the data-owner for special privacy protection against malicious client and cloud collaboration (discussed next) by reducing or abstaining from the vulnerable client's \gls{rm} contribution as a neighbour.

\noindent \textit{Data-Owner.} The data-owner is interested in protecting its \gls{rds} and the \gls{rm}s both from the malicious cloud and the clients.
First, the full \gls{rds} is never exposed, neither to the cloud nor to the clients, instead secret shared \gls{rm}s are provided to the cloud and their respective clients. 
If a malicious client and the cloud collaborate to combine their secret shares, they reveal only the \gls{rm} generated for that particular malicious client.
This substantially limits the information leakage about the full \gls{rds} and if the \gls{rm}s are generated with care, even with several malicious clients collaborating with the cloud, the full \gls{rds} cannot be leaked.\footnote{Alternatively, one can also use matrix factorization to hide the \gls{rm}, however, it requires user interaction and thus was not considered in our use-case.} 
Furthermore, this strategy also protects against malicious clients (without cloud collaboration) relying on query based reconstruction attack, by limiting the client to be able to reconstruct only its own \gls{rm}. In \Cref{subsec:dataowner_processing}, we discuss some of the strategies to generate \gls{rm} while minimizing \gls{rds} leakage.

\noindent \textit{Cloud.} The cloud is interested to protect its proprietary recommendation algorithms from the data-owner and the clients. Due to use of privacy preserving primitives, similar to SANNS, the cloud's internal proprietary algorithms are never exposed.

\begin{figure*}
  \includegraphics[width=10cm]{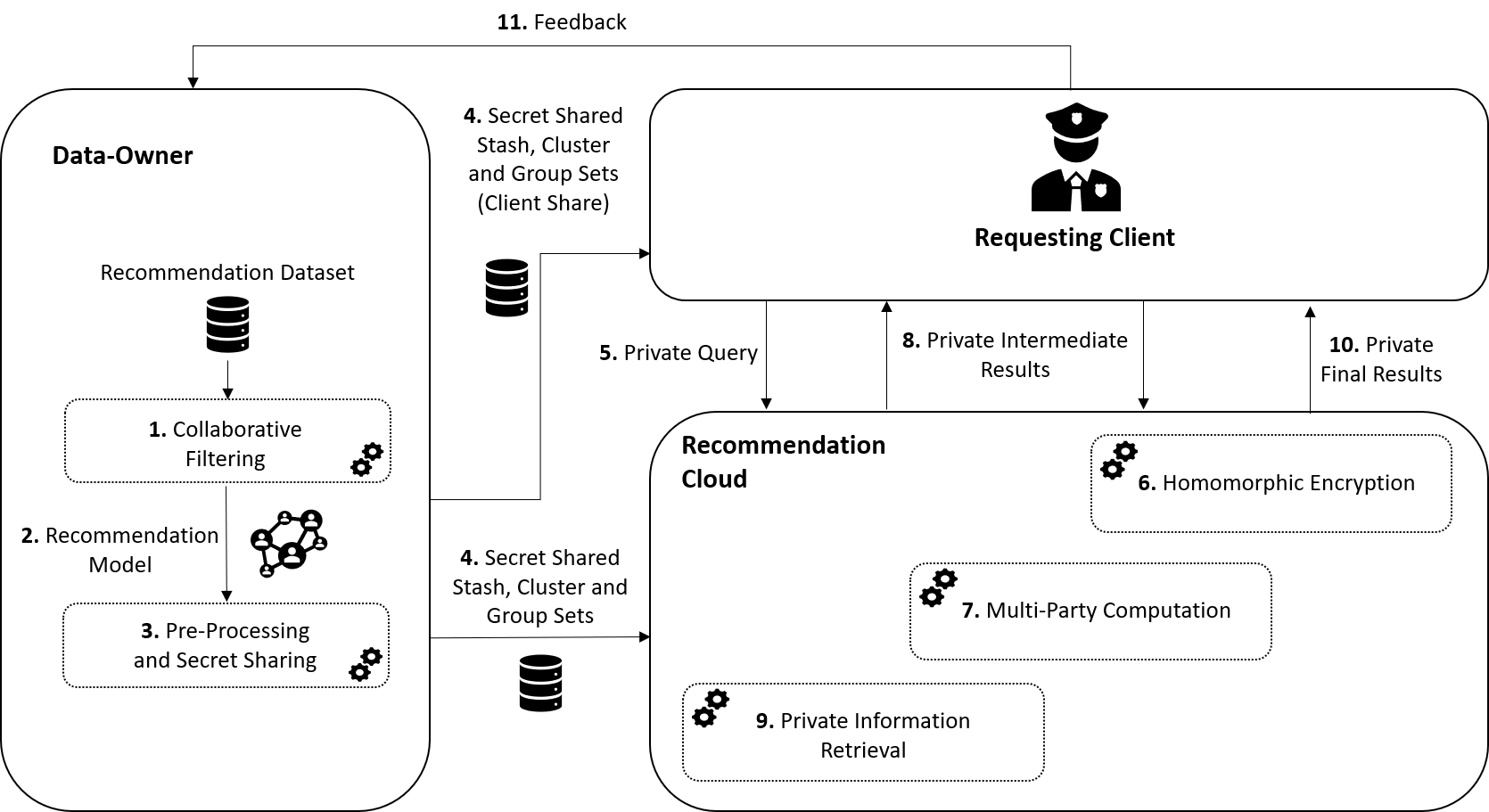}
  \centering
  \caption{Private recommender system protocol at a high level.}
  \label{fig:highlevelprotocol}
\end{figure*}

\subsection{High Level Protocol}
\label{subsection:high_level_protocol}

The collaborative filter algorithm (1) (\Cref{fig:highlevelprotocol}) takes the \gls{rds} as an input and generates a \gls{rm} (2) for each requesting client. The \gls{rm} is pre-processed (3) into stash, cluster and group sets (discussed later) which are then secret shared (4) between the client and the cloud. Similar to SANNS, we use privacy-preserving cryptographic primitives like \gls{he} (6) and \gls{mpc} (7) to guarantee data privacy during all the interactions (5,8,10) between the client and the cloud. The client query (5), intermediate results (8) and the final results (10) are homomorphically encrypted by the client with its secret key, allowing the cloud to compute both on the encrypted client query and the intermediate data before sending the final encrypted results back to the client. \gls{mpc}, more specifically \gls{gc}, enables the client and the cloud to perform private comparisons required to generate the recommendations. Moreover, all the intermediate results are additively secret shared between the client and the server, guaranteeing privacy through information-theoretic security when the client decrypts its homomorphically encrypted intermediate results. \gls{pir} (9) guarantees that a client can privately retrieve any intermediate information from the cloud without revealing its choice, thus protecting from attacks like access pattern analysis. We propose to construct a \gls{pir} using \gls{mpc}-friendly ciphers (low multiplicative complexity) like Kreyvium~\cite{kreyvium} and LowMC~\cite{lowmc}. The security guarantees of the \gls{pir} comes from the security of cipher and \gls{gc} in general. Lastly, after the client has received its recommendation, it can send an optional feedback (11) to the data-owner to improve the recommendation over the time.

\subsection{Data-Owner Processing}
\label{subsec:dataowner_processing}

\noindent \textit{Generating Recommendation Model.} (1,2) We use collaborative filtering (CF)~\cite{herlocker2000explaining} to generate the \gls{rm}s, in particular, we employ a recently proposed adaptation of \gls{cf} called ReuseKNN~\cite{mullner2022reuseknn}, which includes a novel neighborhood selection mechanism to identify a smaller set (\gls{rm}) of $k$ similar neighbors whose ratings can be reused for many different recommendation requests where only this set is exposed or needs to be protected. For $n$ users (targetuser) present in the \gls{rds}, we generate $n$ \gls{rm}s using the ReuseKNN algorithm. During the recommendation process, the cloud selects the appropriate \gls{rm} for the requesting client (targetuser) to generate accurate recommendations. When generating the \gls{rm}, there are two main steps in the ReuseKNN algorithm. The first step computes the rating similarity between the targetuser and the neighbors present in the \gls{rds} using some similarity metric like cosine similarity. The second step estimates a neighbor's reusability score for the targetuser. Please note that Muellner et al. \cite{mullner2022reuseknn} describe several strategies to estimate a neighbor's reusability. In our work, we leverage their personalized neighborhood reuse strategy (Gain) that computes, for how many ratings provided by the targetuser, could a neighbor have covered in the previous recommendation queries. Each neighbor is assigned a final score based on the combined weighted similarity and reusability score. For every item in the \gls{rds} that has not been rated by the targetuser we choose the top-$k$ neighbors who have rated that item and choose their ratings to construct the \gls{rm}. The size of a \gls{rm} can be computed as $|\text{\gls{rm}}| = (m - c_i) \cdot k$ where $m$ is the number of items present in \gls{rds}, $c_i$ is the number of items already rated by the $i^{th}$ client (targetuser) and $k$ is the number of top neighbors who provide the ratings for a particular unseen item. As one may observe, a dataset with more users and fewer items provides better compression rate $|\text{\gls{rm}}|/|\text{\gls{rds}}|$ and also better privacy protection to the \gls{rds}. Increasing $k$ makes the accuracy of the \gls{rm} better, however, it also increases the \gls{rds} exposure if a malicious client collaborates with the cloud. If a particular neighbour is a vulnerable user requiring special protection, our approach allows us to ignore that neighbour and take the ratings of the next suitable neighbour.

\noindent \textit{Pre-Processing Recommendation Model.} (3) Once the \gls{rm} is generated by the data-owner, it needs to be pre-processed for the cloud to generate efficient client recommendations. However, unlike SANNS, instead of pre-processing the \gls{rds}, we pre-process the \gls{rm}s, making the overall \gls{prs} even faster and more scalable than SANNS. Using the K-Means clustering algorithm, the data-owner clusters the \gls{rm} items into clusters. For fair representation, depending on the cluster size, either the clusters are clustered into groups or the items contained in the clusters are moved to a stash set. After the pre-processing phase, the data-owner gets two data sub-sets (stash and cluster) for each \gls{rm}. We refer the reader to SANNS~\cite{sanns} for more details regarding the pre-processing phase.

\noindent \textit{Encrypting Sets.} (4) After preprocessing, once the stash and cluster sets are generated, they must be secret shared between the client and the cloud such that the sets are not exposed to either party without an active collaboration. Using a randomly generated client key, the cloud generates a random stream using the stream cipher Kreyvium and sends it to the data-owner. This random stream is then used to mask (encrypt) the pre-processed sets before sending them to the client. The cloud masks the random client key with its primary secret key and sends the masked key to the client.


\subsection{Cloud Processing}
\label{subsec:cloud_processing}

\noindent \textit{Anonymous \gls{prs}.} (5) Every client has a masked (encrypted) sets which must be unmasked (decrypted) before it can be used as an input to the \gls{prs} algorithm. In order to achieve this the client and the cloud jointly unmask the sets using Kreyvium in \gls{gc} with their respective secret keys before continuing with the \gls{he} KNN computation of the client's query.

\noindent \textit{KNN Computation.} (6,7,8) For computing KNN, we use the simplified Euclidean distance without the square root d$(a,b) = \sum_{i=1}^{n} (a_i-b_i)^2$ as we do not require the actual distance. Since we are computing the distance using \gls{he}, both the client query and the items must be \gls{he} encoded\footnote{We use the BFV scheme supported in the open-source SEAL library due to its fit in our use-case like handling only unsigned integers.}. Among many possible \gls{he} encodings, this work uses the coefficient encoding where the \gls{he} plaintext is represented as a polynomial of degree $d$ with $d+1$ coefficient holding a plain value. Any addition or multiplication is similar to element-wise vector operation in the \gls{he} plaintext polynomial enabling SIMD (Single Instruction, Multiple Data) computation. The client encodes its query with coefficient encoding, encrypts it with its private key and sends it to the cloud. The cloud also encodes its items with the same encoding and computes the distance between the query and the items using \gls{he}. After the distance is computed, the cloud secret shares the distances with the client as intermediate results, which the client decrypts using its \gls{he} key. In the end, the cloud learns nothing about the client's query and both the client and the cloud learn nothing about the intermediate results.

\begin{remark}\footnotesize
    Instead of using KNN, one can choose any other algorithm including neural networks to generate better recommendations. Nonetheless, similar to SANNS, we want to emphasis that KNN is the simplest algorithm that can be efficiently computed with \gls{he} and thus any other algorithm will simply demonstrate a different accuracy-performance trade-off. Since we primarily demonstrate the benefits of combing privacy-aware ML with privacy preserving primitives, exhibiting such trade-off for different algorithms is out of scope for this work. Furthermore, we conjecture that irrespective of the algorithm used, the privacy protection should stay the same if the privacy preserving primitives are used correctly.
\end{remark}

\noindent \textit{KNN Comparison.} (8) After the distance computation, the client and the cloud combine their secret shared distances using \gls{gc} and compare them to find the top-$k$ items with the least distance (most similar to the client query). If the output of the \gls{gc} is the final result (when using stash set), the client learns the result in plain, otherwise (when using cluster sets), the result is again secret shared between the client and the cloud as an intermediate result.

\noindent \textit{Private Information Retrieval.} (9) When the KNN computation is performed on the cluster sets, the client learns the top-$k$ clusters. However, knowing the top-$k$ cluster serves no benefit as the client is interested in the items contained inside top-$k$ clusters. In order to retrieve those items privately from the cloud without revealing the cluster identity, the client uses \gls{pir}. SANNS proposes to use the \gls{fss} based \gls{orom} called Floram~\cite{doerner2017scaling} for \gls{pir}. We alternatively opt for a trivial approach by constructing a \gls{pir} with \gls{mpc}. After privately retrieving the items contained in the top-$k$ cluster, the client and the cloud perform the same computation as they did to find the stash top-$k$ items. For more details on the cloud processing, we refer reader to SANNS~\cite{sanns}.

\noindent \textit{Improving \gls{prs}.} (11) Improving the recommendation quality over time is key to any successful \gls{rs}. In our approach, after a client has viewed its recommendation, it can send a feedback to the data-owner who updates the \gls{rds} and the \gls{rm} for that client (targetuser) and sends the new secret shared pre-processed sets to the cloud.

\section{Experiments}

\noindent \textit{Environment.} We used two different environments to host our \gls{prs} cloud. For the standard CPU benchmarking, we use Intel i5-8250U, 24 GB of RAM running Ubuntu 20.04.3. The \gls{soc} benchmarking uses Intel Celeron N4100 with 4 GB of RAM running Ubuntu 20.04.3. We perform the network benchmarks on LAN settings and provide traffic analysis for the same to estimate on other network conditions.

\noindent \textit{Datasets.} We use two standard benchmarking datasets, namely, the MovieLens\footnote{\url{https://grouplens.org/datasets/movielens}} (100k, 1M, 10M and 20M) where we have up to 20 million ratings with 27278 movies rated by 138493 users and the Netflix\footnote{\url{https://www.kaggle.com/datasets/netflix-inc/netflix-prize-data}} 100M dataset containing 100 million ratings with 17770 movies rated by 480189 users. Each dataset contains \textit{userid}, \textit{movieid} and \textit{ratings} attributes where the ratings range between 1 and 5 with an interval of 0.5 and 1 for the MovieLens and Netflix dataset respectively.

\noindent \textit{Accuracy.} We achieve a 9-out-of-10 correct KNN recommendation accuracy\footnote{The accuracy of a private algorithm should match to that of a plain non-private algorithm as similar computations are performed in both approaches without loss in accuracy. We emphasize that the slight drop in accuracy, discussed above, is due to the stash and cluster optimization, similar to SANNS, leading to a runtime/memory-accuracy trade-off.}, similar to SANNS, for all the datasets. In comparison to the prior time-memory-accuracy trade-offs discussed in SANNS, ReuseKNN provides a broader and more flexible trade-off between time-memory-accuracy and privacy. For example, when decreasing the $k$-nearest neighbour size for each item from 3 to 2, the \gls{rm} size reduces by 33\%. Due to the linear time and memory complexity of our \gls{prs} with respect to the \gls{rm} size, we witness a similar decrease both in time and memory requirement, however, the accuracy drops to 7-out-of-10 as a trade-off.


\begin{table}[ht]
\setlength{\tabcolsep}{6pt}
  \centering
  \caption{Time-Memory improvement when using recommendation model approach in comparison to the previous approach as taken in SANNS.}
  \label{tab:timememoryimprovement}
  \begin{tabular}{lrrr}
  \toprule
  \multicolumn{1}{c}{Dataset} & \multicolumn{1}{c}{Model Size} &  \multicolumn{1}{c}{Time Improvement} &  \multicolumn{1}{c}{Memory Improvement} \\
  \midrule
    MovieLens 100K & 137\,KB & $\times10$ & $\times7$ \\
    MovieLens 1M & 322\,KB  & $\times63$ & $\times42$  \\
    MovieLens 10M & 740\,KB & $\times271$ & $\times205$ \\
    MovieLens 20M & 1638\,KB & $\times361$ & $\times222$ \\
    Netflix 100M & 1249\,KB & $\times2235$ & $\times1136$ \\
  \bottomrule
  \end{tabular}
\end{table}


\noindent \textit{Time-Memory-Privacy Improvement.} Here we present the performance comparison when our \gls{prs} runs with/without the proposed optimization. For MovieLens 100k and 1 million datasets, we improve by an order of magnitude (except memory for 100K), whereas, for 10M and 20M datasets, it improves by two orders of magnitude for both time and memory. Lastly, when using the ReuseKNN optimization for the Netflix 100M dataset, we witness three orders of magnitude improvement compared to not using the ReuseKNN optimization. Refer to \Cref{tab:timememoryimprovement} for more details. The ReuseKNN optimization enables the data-owner to reduce the average user data exposure by up to 93\% 
as a part of the privacy-accuracy trade-off. We measure this amount by calculating the average difference in 
the number of items rated by a neighbour used (exposed contribution) for generating targetuser recommendations. By tweaking the neighbour contribution when generating the \gls{rm}s, our approach enables us to protect the special vulnerable users.

\noindent \textit{Standard and SOC Benchmark.} Using our SOC device as a cloud, processing a single \gls{prs} query with the 10M MovieLens dataset takes under a minute. Similarly, for the 100M Netflix dataset, a private recommendation is generated in around 100 seconds, further demonstrating the efficiency of our proposal even on power constrained device. For comparison, we additionally provide the time required by a standard CPU when computing the same. On average the standard CPU is x3 to x6 times faster than the SOC device. Refer to \Cref{tab:singlethreadtimememoryperformance} for more details.


\begin{remark}\footnotesize
    At the time of writing this paper, the implementation of SANNS was not publicly available, due to which, any direct comparison between SANNS and our work is not possible. Moreover, due to our work solely relying on privacy preserving cryptographic primitives when generating recommendations, we limit ourselves from comparing with other proposals relying on non-rigorous privacy techniques like trusted party, secure hardware, among others, as they naturally enjoy a better efficiency when considering a much weaker privacy assumption.
\end{remark}


\begin{table}[ht]
\setlength{\tabcolsep}{6pt}
  \centering
  \caption{Time-Memory performance of a single query with our optimized approach. For the Netflix 100M benchmark, additional 2 GB swapping (using built-in eMMC storage) was used for the SOC device to compute the private recommendation.}
  \label{tab:singlethreadtimememoryperformance}
  \begin{tabular}{lrrrr}
  \toprule
  \multicolumn{1}{c}{Dataset} & \multicolumn{1}{c}{SOC Time} &  \multicolumn{1}{c}{Std. CPU Time} &  \multicolumn{1}{c}{Memory} &  \multicolumn{1}{c}{Network} \\
  \midrule
   MovieLens 100K & $15\,s$ & $4\,s$ & $1.3$\,GB & 32\,MB \\
   MovieLens 1M & $17\,s$ & $6\,s$ & $1.9$\,GB & 66\,MB \\
   MovieLens 10M & $53\,s$ & $14\,s$ & $2.8$\,GB & 106\,MB \\
   Netflix 100M & $103\,s$ & $17\,s$ & $4.2$\,GB & 146\,MB \\
  \bottomrule
  \end{tabular}
\end{table}


\section{Further Discussion}
\label{sec:furtherdiscussions}

Our approach assumes that the data-owner owns the \gls{rds} and wants to protect it. However, protecting the privacy of the users who are providing their data to the data-owner remains an open problem, especially, when the data-owner cannot be trusted. Several works 
\cite{zhao2015privacy,melis2015efficient,chen2021differentgially} have discussed numerous techniques to privately perform \gls{cf}, however, all of them rely on trusted-party or active user communication, in addition to having scalability issues, leaking recommendation models, among others. Trivially, one can also use \gls{mpc} as a solution but we conjecture that such an approach is not scalable due to the involved computation and communication overhead. Another standard technique is data perturbation \cite{polat2003privacy}, where user data is protected by adding some structured noise. This work heuristically explored the possibility of such a technique with a rating domain $\{0 \text{ (bad)},1\text{ (dislike)},2\text{ (unrated)},3\text{ (liked)},4\text{ (good)}\}$. The unrated value $2$ is placed in the middle to reduce the overall impact on the accuracy when perturbing the ratings. For perturbation, the noise was drawn from a zero-centered ($\mu = 0$) Laplace distribution Lap($\mu,b$) (rounded into the rating domain), where $b = 0.7$ was found to be the most suitable in our use-case.\footnote{We naively selected a small $b$ and raised it until the accuracy suffered.} Unfortunately, the above privacy guarantees are based on heuristics and thus a shift towards a more rigorous privacy guarantee like \gls{dp}~\cite{dwork2006calibrating}, especially the notion of \gls{ldp}~\cite{cormode2018privacy}, actively used in the industry~\cite{bittau2017prochlo,ding2017collecting,shen2016epicrec,appleldp} and research~\cite{bampoulidis2020privately,chen2021differentgially} would be desirable. We leave the incorporation of \gls{dp} for future work.

\section{Conclusion}

Using cryptographic primitives for constructing \gls{prs} was so far discouraging due to the high cost and little flexibility regarding time-memory-accuracy-privacy trade-offs. 
With this work, we proposed a significantly more scalable and efficient \gls{prs} by combining cryptographic primitives with machine learning algorithms for the first time. 
We also demonstrated that adopting our new approach provides improved privacy, especially, for vulnerable users who may require additional protection.
Moreover, our proposed model allows us to describe a more realistic scenario where not only the client query and recommendation is protected from the cloud but also the proprietary \gls{rds} of the data-owner is protected both from the client and the cloud.
Finally, we release our solution as open-source with the intention of faster adoption and integration of \gls{prs} into the existing non-private recommendation system infrastructure for more secure and private access to news, social media, academic papers, or entertainment artifacts.

\subsection*{Acknowledgements}
This work received funding from the Horizon Europe Program under grant agreement number 101096435 (“CONFIDENTIAL6G"), and is supported by the "DDAI" COMET Module managed by Austrian Research Promotion Agency (FFG).

%
%
%
\ifllncs
\bibliographystyle{splncs04}
\fi
\ifiacrtrans
\bibliographystyle{alpha}
\fi

\bibliography{references}

\appendix

\end{document}